\documentclass[preprint,amsmath,amssymb,article,pra]{revtex4-1}
\usepackage{dcolumn}
\usepackage{bm}
\usepackage{amsfonts}
\usepackage{physics}
\usepackage{amsmath}
\usepackage{amssymb}
\usepackage{float}
\usepackage{graphicx}
\usepackage{subfigure}
\usepackage{epstopdf}
\usepackage{hyperref}
\usepackage{placeins}
\usepackage[dvipsnames]{xcolor}

\begin{document}
\title{Robustness of entanglement in a non-Hermitian cavity-optomechanical system even away from exceptional points}

\author{Jia-Jia Wang$^{1}$}
\author{Yu-Hong He$^{1}$}
\author{Chang-Geng Liao$^{1,2}$}
\thanks{cgliao@zjgsu.edu.cn}
\author{Rong-Xin Chen$^{3}$}
\thanks{chenrxas@163.com}
\author{Jacob A. Dunningham$^{2}$}
\thanks{J.Dunningham@sussex.ac.uk}

\affiliation{$^{1}$ School of Information and Electronic Engineering (Sussex Artificial Intelligence Institute), Zhejiang Gongshang University, Hangzhou, 310018, China}
\affiliation{$^{2}$ Department of Physics and Astronomy, University of Sussex, Brighton BN1 9QH, United Kingdom}
\affiliation{$^{3}$ School of Mechanical and Electrical Engineering, Longyan University, Longyan, 364012, China}


\begin{abstract}
Quantum physics can be extended into the complex domain by considering non-Hermitian Hamiltonians that are $\mathcal{PT}$-symmetric. These exhibit exceptional points (EPs) where the eigenspectrum changes from purely real to purely imaginary values and have useful properties enabling applications such as accelerated entanglement generation and the delay of the sudden death of entanglement in noisy systems. An interesting question is whether similar beneficial effects can be achieved away from EPs, since this would extend the available parameter space and make experiments more accessible. We investigate this by considering a $\mathcal{PT}$-symmetric optomechanical system but also consider what happens when two-mode squeezing interactions are included, taking us into the pseudo-Hermitian regime. The addition of squeezing is motivated by an attempt to extend the lifetime of the system's entanglement. While this does not prove to be the case, rich dynamics are nonetheless observed in both the pseudo-Hermitian and $\mathcal{PT}$-symmetric systems, including the sudden death and revival of entanglement under certain conditions. In both cases, we find that the sudden disappearance of entanglement can be mitigated at EPs, and also show that the revival of entanglement is quite robust to thermal noise in a group of parameters away from the EPs. This investigation extends our understanding of non-Hermitian systems and opens a new perspective for the development of quantum devices in non-Hermitian systems even away from EPs.
\end{abstract}

\maketitle
\section{Introduction}
In 1998 Bender and Boettcher extended standard quantum theory into the complex domain by considering Hamiltonians that need not be Hermitian but rather exhibit the property of being invariant when subjected to both $\mathcal{P}$ (parity) and $\mathcal{T}$ (time) inversions \cite{bender1998real,bender2002complex}. These $\mathcal{PT}$-symmetric Hamiltonians have the property that they commute with the $\mathcal{PT}$ operator, i.e. $[H, \mathcal{PT}] = 0$ \cite{el2018non}. When $\mathcal{PT}$-symmetry holds, the Hamiltonian has a real spectrum (despite being non-Hermitian) and when the symmetry is broken, complex-conjugate pairs of eigenvalues emerge \cite{mostafazadeh2002pseudo}. The parameter values where these transitions occur are called exceptional points (EPs) and are where nontrivial degeneracies allow eigenvalues and eigenstates to coalesce. This characteristic has been experimentally demonstrated in a range of  systems \cite{dembowski2001experimental,ruter2010observation,regensburger2012parity}, including diverse quantum platforms such as exciton-polaritons \cite{gao2015observation}, nitrogen-vacancy centers  \cite{wu2019observation,liu2021dynamically}, trapped ions  \cite{ding2021experimental,wang2021observation}, ultracold atoms \cite{li2019observation}, and superconducting circuits  \cite{naghiloo2019quantum,abbasi2022topological,han2023exceptional}, offering new possibilities for quantum applications including wave transport control  \cite{feng2011nonreciprocal,doppler2016dynamically,choi2017extremely,zhang2018dynamically,liu2018unidirectional}, laser emission management \cite{peng2014loss,hodaei2014parity,brandstetter2014reversing,peng2016chiral,wong2016lasing}, topological energy transfer \cite{xu2016topological}, and enhanced sensing \cite{wiersig2014enhancing,liu2016metrology,chen2017exceptional,hodaei2017enhanced,yu2020experimental}. The phenomenon of linewidth narrowing can be induced by $\mathcal{PT}$-symmetric feedback \cite{tang2023pt} and can further optimize and control the performance of these quantum systems. Furthermore, when non-Hermitian behavior is combined with topology, it is possible to achieve robustness and flexible control of the system at the same time \cite{gao2015observation,abbasi2022topological,qian2024probing}. Recent work \cite{han2023exceptional} has explored the question of when non-Hermitian physics departs from classical physics and has a truly quantum signature.

In 2002, Mostafazadeh extended the idea $\mathcal{PT}$-symmetry by introducing the concept of pseudo-Hermitian systems~\cite{mostafazadeh2002pseudo}, in which a Hamiltonian \( H \) is defined to be pseudo-Hermitian with respect to the parameter \(\eta\) by $H^\dagger = \eta H \eta^{-1}$, where \(\eta\) is an invertible operator.
When \(\eta\)=1, this reduces to the standard Hermiticity condition, \( H^\dagger = H \). If the Hamiltonian satisfies the condition $H^\dagger = P H P^{-1}$, then the operator \( H \) is called $P$-pseudo-Hermitian. Therefore, all Hermitian systems are special cases of pseudo-Hermitian systems.  While pseudo-Hermiticity and $\mathcal{PT}$-symmetry share some similarities, such as the possibility of real energy spectra, they differ in several properties and potential applications. The presence of non-Hermitian effects was previously regarded as a problem and efforts were made to suppress them. Nevertheless, the realization that non-Hermitian effects may be an extension of quantum mechanics has prompted a surge of interest in this field \cite{pinske2019holonomic,deguchi2009level,simeonov2016dynamical}. Like $\mathcal{PT}$-symmetry, pseudo-Hermiticity has numerous applications, including the description of spinor fields in gravitational wave Kerr fields \cite{gorbatenko2013stationary}, optical microspiral cavities \cite{wiersig2008asymmetric}, and micro-cavities perturbed by particles \cite{wiersig2011structure}. It describes the weak backscattering between counter-propagating travelling waves in a general open quantum system \cite{wiersig2014chiral} and models light propagation in a perturbed medium \cite{Yariv1973,boyd2007nonlinear}. It has also been applied in fields as diverse as quantum cosmology, magnetohydrodynamics and quantum chaos \cite{mostafazadeh2010pseudo}. The unconventional behavior in pseudo-Hermitian systems, particularly around EPs, has also led to unique and unintuitive phenomena, such as geometric phases \cite{berry1984quantal} and non-orthogonal eigenstates \cite{brody2013biorthogonal}. These phenomena not only expand our understanding of quantum systems but also provide potential new avenues and tools for the development of quantum technologies and devices.

Quantum entanglement \cite{horodecki2009quantum} is special kind of nonclassical correlation and is the key resource in many quantum technologies. Being able to preserve entanglement is therefore important, but it is inevitable that entanglement will interact with its surroundings affecting its lifetime. One of these detrimental manifestations is entanglement sudden death (ESD) \cite{yu2009sudden}, where quantum entanglement suddenly and completely disappears within a finite time, in contrast to the gradual decay that might be expected. Entangled states can be generated by two-mode squeezing \cite{ou1992realization,reid1988quantum,caves1985new,schumaker1985new} and exceptional-point-enabled entanglement behaviors have been experimentally demonstrated in a superconducting circuit platform \cite{han2023exceptional}. Additionally, the precise control of the gain and loss in optical cavities has been used to realize $\mathcal{PT}$-symmetry in these systems ~\cite{regensburger2012parity}, making  $\mathcal{PT}$-symmetric cavity optomechanical systems possible~\cite{xu2016topological}. Theoretical work has shown  that at EPs, the sudden death of entanglement can be significantly delayed in a $\mathcal{PT}$-symmetric cavity optomechanical system \cite{chakraborty2019delayed}.

This paper sets out to investigate the dynamics and robustness of entanglement in a $\mathcal{PT}$-symmetric system and a closely-related pseudo-Hermitian one, since the unconventional evolution in such systems impacts the generation and decay of entangled states. The pseudo-Hermitian system we consider is obtained by including two-mode squeezing interactions into the $\mathcal{PT}$-symmetric optomechanical system considered in ~\cite{chakraborty2019delayed} and we closely follow the formalism of that work.  The addition of squeezing is motivated by an attempt to extend the lifetime of the system's entanglement. Surprisingly, we find that the squeezing does not generally preserve the initial entanglement, and the dynamics of entanglement are hardly different between our  pseudo-Hermitian and $\mathcal{PT}$-symmetric systems. In both, the sudden disappearance of entanglement can be mitigated at EPs, and the revival of entanglement is quite robust to thermal noise in a group of parameters away from EPs. This is a useful feature not found in related work~\cite{chakraborty2019delayed} that may enable applications, such as enhanced sensing, to be achieved away from EPs.

The paper is organized as follows. In Section \ref{sec:Systems}, the model of the pseudo-Hermitian optomechanical system is introduced. We then study the behavior of entanglement near EPs and analyze the impact of thermal noise on the evolution of two-mode entanglement in binary systems in Section \ref{sec:Entanglement in binary systems}. In Section \ref{sec:Entanglement in ternary systems}, the results are extended to ternary systems. Finally, conclusions are given in Section \ref{sec:Conclusions}.

\section{The model}\label{sec:Systems}

We consider a generic cavity optomechanical system, which is composed of a single cavity mode and a mechanical mode with frequencies \(\omega_\text{c}\) and \(\omega_\text{m}\), respectively. In the rotating frame with respect to laser frequency \(\omega_\text{l}\) the Hamiltonian of the system with \(\hbar = 1\) is \cite{aspelmeyer2014cavity}
\begin{align}
H = \Delta_0 a^\dagger a + \omega_\text{m} b^\dagger b - g a^\dagger a (b^\dagger + b) + E_0 (a^\dagger + a),
\end{align}
where \( \Delta_{0} = \omega_\text{c} - \omega_\text{l} \) is the cavity detuning; \( a \) (\( a^\dagger \)) and  \( b \) (\( b^\dagger \)) are the annihilation (creation) operators of the cavity field and the mechanical resonator respectively; \( g \) is the single-photon coupling strength; and \( E_{0} \) is the driving amplitude. Taking the fluctuation and dissipation processes into account, the evolution of the system is described by the nonlinear quantum Langevin equations (QLEs):
\begin{subequations}
  \begin{align}
\dot{a} &= -(i\Delta_0+\kappa/2)a + iga( b^\dagger+b) -iE_0+\sqrt{\kappa}a^{\text{in}}, \\
\dot{b} &= -(i\omega_\text{m}+\gamma/2)b + iga^\dagger a +\sqrt{\gamma}b^{\text{in}}.
\end{align}
\end{subequations}
Here, $\kappa$ is the cavity decay rate, $\gamma$ is the mechanical damping rate, and $a^{\text{in}}$ and $b^{\text{in}}$ are independent zero-mean vacuum input noise operators, which satisfy the correlation functions
\begin{subequations}
 \begin{align}
\langle a_{\text{in}}(t) a_{\text{in}}^\dagger(t') \rangle &= \delta(t - t'),  \\
\langle b_{\text{in}}^\dagger(t) b_{\text{in}}(t') \rangle &= n_{\text{th}} \delta(t - t'), \\
\langle b_{\text{in}}(t) b_{\text{in}}^\dagger(t') \rangle &= (n_{\text{th}} + 1) \delta(t - t'),
\end{align}
\end{subequations}
where \( n_{\text{th}}  = \left[ \exp(\hbar \omega_\text{m}/k_\text{B} T) - 1 \right]^{-1}\) is the mean thermal phonon number at temperature \( T \) and \( k_\text{B} \) is the Boltzmann constant.

For strongly driven cavities, we can follow the standard linearization procedure \cite{vitali2007optomechanical} where we introduce a $c$-number for the steady state value of each classical field ($\alpha$ and $\beta$) and add to them zero-mean quantum fluctuations given by the $a$ and $b$ operators respectively. This gives rise to equations that can be solved for $\alpha$ and $\beta$ and the linearized QLEs for the quantum fluctuations \cite{chakraborty2019delayed}. Next we introduce  two slowly varying quantum fluctuation operators \( \tilde{a} = a e^{i\Delta t} \) and \( \tilde{b} = b e^{i\omega_\text{m} t} \) as well as noise operators \( \tilde{a}^{\text{in}} = a^{\text{in}} e^{i\Delta t} \) and \( \tilde{b}^{\text{in}} = b^{\text{in}} e^{i\omega_\text{m} t} \), where $\Delta = \Delta_0 -2g{\rm Re}(\beta)$ is the effective cavity detuning. The linearized QLEs can then be written as
\begin{subequations}
\begin{align}
\dot{\tilde{a}} &= -\frac{\kappa}{2} \tilde{a} + i G (\tilde{b}^\dagger e^{i(\Delta + \omega_\text{m})t} + \tilde{b} e^{i(\Delta - \omega_\text{m})t}) + \sqrt{\kappa} \tilde{a}_{\text{in}},\ \\
\dot{\tilde{b}} &= -\frac{\gamma}{2} \tilde{b} + i G (\tilde{a}^\dagger e^{i(\omega_\text{m} + \Delta)t} + \tilde{a} e^{i(\omega_\text{m} - \Delta)t}) + \sqrt{\gamma} \tilde{b}_{\text{in}} \
\end{align}
\end{subequations}
with effective coupling strength $G=g|\alpha|$. For $\Delta = -\omega_\text{m}$ (red-detuned regime), we can neglect the fast oscillating terms under the rotating-wave approximation (RWA) to get
\begin{subequations}
\begin{align}
\dot{\tilde{a}} &= -\frac{\kappa}{2} \tilde{a} + iG \tilde{b}^{\dagger} + \sqrt{\kappa} \tilde{a}^{\text{in}},\ \\
\dot{\tilde{b}} &= -\frac{\gamma}{2} \tilde{b} + iG \tilde{a}^{\dagger} + \sqrt{\gamma} \tilde{b}^{\text{in}}.\
\end{align}
\end{subequations}

When the cavity decay rate is much greater than the effective optomechanical coupling strength and mechanical damping rate, i.e., $\kappa \gg \{ G, \gamma \}$, we can adiabatically eliminate the cavity field to give
\begin{align}
    \tilde{a} = i \frac{2G}{\kappa} \tilde{b}^{\dagger} + \frac{2}{\sqrt{\kappa}} \tilde{a}^{\text{in}}.
\end{align}
The effective dynamics equation of the mechanical resonator with gain is finally obtained by substituting Eq.(6) into Eq.(5b)
\begin{align}
\dot{\tilde{b}} = \left( \frac{\Gamma}{2} - \frac{\gamma}{2} \right) \tilde{b} + i \sqrt{\Gamma} \tilde{a}^{\text{in} \dagger} + \sqrt{\gamma} \tilde{b}^{\text{in}},
\end{align}
where $\Gamma = 4G^2/\kappa$ is the optomechanically-induced gain in the mechanical resonator. For $\Delta = \omega_\text{m}$ (blue-detuned regime), a similar procedure yields
\begin{align}
\dot{\tilde{b}} = -\left( \frac{\Gamma}{2} + \frac{\gamma}{2} \right) \tilde{b} + i \sqrt{\Gamma} \tilde{a}^{\text{in} \dagger} + \sqrt{\gamma} \tilde{b}^{\text{in}}.
\end{align}

\begin{figure}
    \centering
    \includegraphics[width=0.5\linewidth]{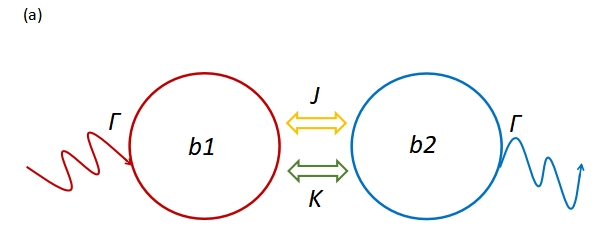}
    \includegraphics[width=0.5\linewidth]{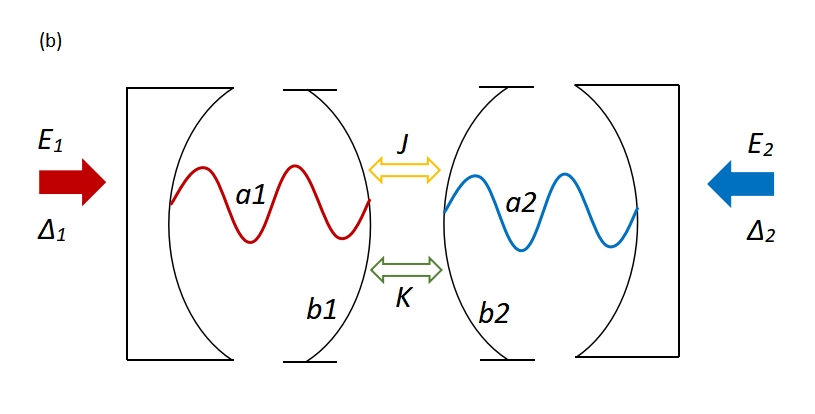}

    \caption{ (a) The pseudo-Hermitian system consisting of two interacting mechanical oscillators with optomechanically induced gain and loss. (b) Scheme for engineering mechanical gain and loss in an optomechanical system where the two mechanical oscillators are coupled with each other though both the two-mode squeezing and beam-splitter-like interaction simultaneously. One cavity is driven in the red-detuned regime and the other in the blue-detuned regime. }
        \label{fig:fig1}
\end{figure}

Chakraborty and Sarma \cite{chakraborty2019delayed} constructed a $\mathcal{PT}$-symmetric optomechanical system by coupling mechanical resonators via a beam-splitter-like interaction, i.e. \( H = -J(b_1^\dagger b_2 + b_1 b_2^\dagger)\) and investigated the dynamics of the system near EPs in the context of ESD. Here we additionally incorporate two-mode squeezing interactions, i.e. \( H = K(b_1^\dagger b_2^\dagger + b_1 b_2) \) to see whether this preserves entanglement for longer. The interaction between two mechanical resonators could be achieved, for example, by the Coulomb interaction \cite{chen2015dissipation,hensinger2005ion,zhang2012precision} or with the assistance of the cavity mode, generating a two-mode squeezing
state \cite{woolley2014two,wang2013reservoir,ockeloen2018stabilized,tan2013dissipation}. This squeezing term leads to a pseudo-Hermitian optomechanical system with effective dynamical equations
\begin{subequations}
\begin{align}
\dot{b}_1 &= \left( \frac{\Gamma}{2} - \frac{\gamma}{2} \right)b_1 + iJb_2 -iKb_2^\dagger + i\sqrt{\Gamma}a_1^{\text{in}\dagger} + \sqrt{\gamma}b_1^{\text{in}},  \\
\dot{b}_2 &= -\left( \frac{\Gamma}{2} + \frac{\gamma}{2} \right)b_2 + iJb_1 -iKb_1^\dagger+i \sqrt{\Gamma}a_2^{\text{in}\dagger} + \sqrt{\gamma}b_2^{\text{in}},
\end{align}
\end{subequations}
where we have dropped the tildes for notational simplicity.
Each resonator is characterized by a frequency (damping rate) \( \omega_\text{m} \) (\( \gamma \)) and has mechanical couplings with strengths \( J \) and \( K \), \( b_1 (b_1^\dagger) \) and \( b_2 (b_2^\dagger) \) are the annihilation (creation) operators of the gain and loss mechanical oscillators respectively, and \( \Gamma \) is the optomechanically-induced effective gain or loss rate. The operator \( a_j^\text{in} \) (\( b_j^\text{in} \)) represents the noise acting on the $j\text{th}$ cavity field (mechanical resonator). A schematic illustration of the pseudo-Hermitian system  and its optomechanical realization are shown in Fig.~\ref{fig:fig1}.

\section{Entanglement in Binary Systems}\label{sec:Entanglement in binary systems}
When the effective optomechanical coupling, $\Gamma$, is much stronger than the intrinsic mechanical damping $\gamma$, we can simplify our theoretical analysis by ignoring $\gamma$. However, we retain a small finite damping $\gamma = 10^{-3}$ in all the numerical simulations in this paper. By neglecting the quantum noise, we can redefine Eq.(9a) and (9b) as
\begin{equation}
\dot{u}(t) = -iH_2 u(t),
\end{equation}
where $u^T(t) = (p_1(t), q_1(t), p_2(t), q_2(t))$ is the state vector and the quadrature operators are defined as $q_j \equiv (b_j + b_j^\dagger) /\sqrt{2}$, $p_j \equiv  (b_j - b_j^\dagger) /i\sqrt{2}$ (with $j = 1, 2$). The non-Hermitian Hamiltonian is given by
\begin{equation}
H_2 = i \begin{pmatrix}
    \frac{\Gamma}{2} & 0 & 0 & -(J+K) \\
    0 & \frac{\Gamma}{2} & J-K & 0 \\
    0 & -(J+K) & -\frac{\Gamma}{2} & 0 \\
    J-K & 0 & 0 & -\frac{\Gamma}{2}
\end{pmatrix}.
\end{equation}
It is easy to show that the Hamiltonian is not $\mathcal{PT}$-symmetric, i.e., \( [\mathcal{PT}, H_2] \neq 0 \), but is pseudo-Hermitian as we can construct the parameter $\eta=S_x\otimes S_y$ to meet the definition of pseudo-Hermitian $H^\dagger = \eta H \eta^{-1}$. Since pseudo-Hermitian systems can also exhibit non-Hermitian features, they can likewise have EPs in parameter space.

The eigenfrequencies, \( \omega_\pm \), of $H_2$ are given by
\begin{equation}\label{eigenfrequencies}
\omega_\pm = \pm \frac{\sqrt{4J^2 - 4K^2 - \Gamma^2}}{2},
\end{equation}
where the real and imaginary parts correspond
to the effective frequency and dissipation of the mechanical
resonator, respectively.
\begin{figure}
    \centering
    \includegraphics[width=0.8\linewidth]{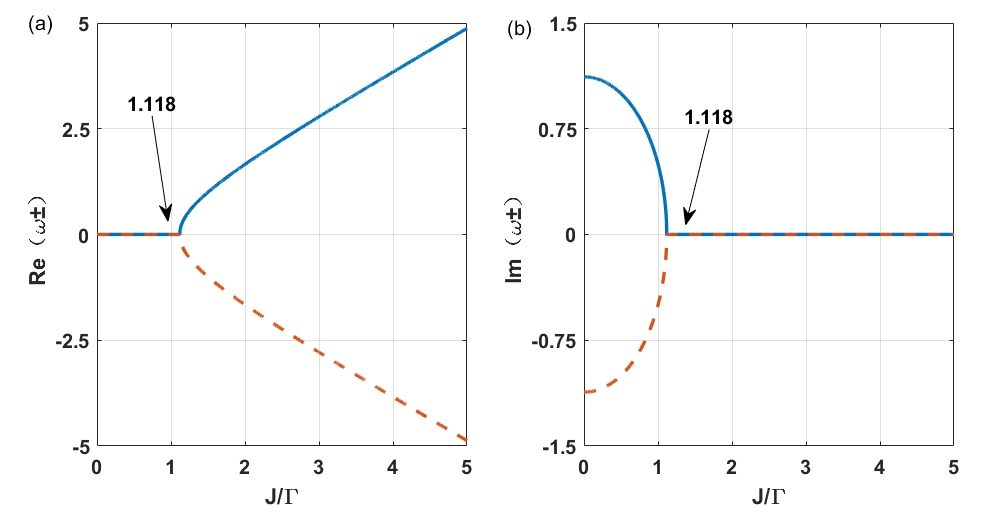}

    \caption{(a) The real and (b) imaginary parts of the eigenfrequencies \( \omega_\pm \) as a function of the ratio of coupling coefficients \( J/\Gamma \) when \( K/\Gamma  = 1 \). The exceptional point appears at \( J/\Gamma = 1.118 \).}
    \label{fig:fig2}
\end{figure}
Fig.~\ref{fig:fig2} shows the real and imaginary parts of the eigenfrequencies \( \omega_\pm \) as functions of the ratio of coupling strengths \( J/\Gamma \) when the ratio \( K/\Gamma = 1 \) is fixed. We see that there exist two distinct phases similar to those occurring with $\mathcal{PT}$ symmetry: one is a pseudo-Hermitian symmetric phase with real eigenvalues and the other is a broken pseudo-Hermitian symmetric phase with purely imaginary eigenvalues~\cite{mostafazadeh2010pseudo}.  In this case, although the Hamiltonian is not Hermitian, it can be related to its adjoint operator through a similarity transformation. This symmetry allows the eigenvalues of the Hamiltonian to be real. In the pseudo-Hermitian symmetric phase, when \(J > 1.118\Gamma\), all eigenvalues of the Hamiltonian are real, indicating that the system is physically similar to a Hermitian system. Conversely, when \(J < 1.118\Gamma\), in the broken pseudo-Hermitian symmetry phase, some of the eigenvalues of the Hamiltonian become complex numbers, with their imaginary parts representing gains or losses in the system. This signifies that the system is no longer stable, and its dynamic behavior will exhibit oscillatory, decaying, or growing patterns. The EPs at \(J = 1.118\Gamma\) in Fig.~\ref{fig:fig2} marks the transition from the pseudo-Hermitian symmetric phase to the broken phase.

Next, we include quantum noise in the model. Eq.~(9) can be rewritten as $\dot{u}(t) = Au(t) + n(t)$, where the drift matrix \( A = -iH_2 \), and the matrix of corresponding noises is given by
\begin{equation}
n^T(t) = \left( \sqrt{\Gamma} Y_1^{\text{in}} + \sqrt{\gamma} Q_1^{\text{in}}, \sqrt{\Gamma} X_1^{\text{in}} + \sqrt{\gamma} P_1^{\text{in}}, -\sqrt{\Gamma} Y_2^{\text{in}} + \sqrt{\gamma} Q_2^{\text{in}}, \sqrt{\Gamma} X_2^{\text{in}} + \sqrt{\gamma} P_2^{\text{in}} \right).
\end{equation}
Here, $X_{j}^{\text{in}} \equiv ({a_{j} + a_{j}^\dagger})/{\sqrt{2}},  Y_{j}^{\text{in}} \equiv ({a_{j} - a_{j}^\dagger})/{i\sqrt{2}}$, and $Q_{j}^{\text{in}} \equiv ({b_{j} + b_{j}^\dagger})/{\sqrt{2}},  P_{j}^{\text{in}} \equiv ({b_{j} - b_{j}^\dagger})/{i\sqrt{2}}$  (with $j = 1, 2$).
Since the linearized Hamiltonian ensures that the evolved states retain their initial Gaussian characteristics, we can fully describe the system by the covariance matrix~\cite{braunstein2005quantum}. This allows us to derive the following linear differential equation
\begingroup
\setlength{\abovedisplayskip}{4pt}
\setlength{\belowdisplayskip}{4pt}
\begin{equation}
\dot{V}(t) = AV(t) + V(t)A^T + D
\end{equation}
\endgroup
where \( D = \left[ {\Gamma}/{2} + \gamma \left( {1}/{2} + n_{\text{th}} \right) \right] \mathrm{diag}(1, 1, 1, 1) \) and each element of the matrix $V$ is defined  by $V_{ij}(t) = \{\left\langle u_i(t) \, u_j(t) \right\rangle +\left\langle u_j(t) \, u_i(t) \right\rangle\}/{2}$. We are interested in the behavior of entanglement in this system. The initial state is taken to be a two-mode squeezed state \( |z\rangle = e^{r (b_1^\dagger b_2^\dagger - b_1 b_2)} |0, 0\rangle \), with \( r \) being the squeezing parameter~\cite{ou1992realization,reid1988quantum,caves1985new,schumaker1985new}. Entangled optomechanical oscillators have already been realized~\cite{ockeloen2018stabilized,riedinger2018remote,marinkovic2018optomechanical}. For convenience, we ignore quantum noise and mechanical damping \(\gamma\) in deriving the following analytical solutions for each variance-related term in \(V(t)\) (noise will be reintroduced in our numerical calculations):
\begin{subequations}
\begin{align}
V_{11} \equiv& \langle q_1^2 \rangle = \frac{\cosh(2r)}{2} \left( \frac{J^2 + JK}{\omega_+^2} - \frac{4JK + 4K^2 + \Gamma^2}{4\omega_+^2} \cos(2\omega_+ t) + \frac{\Gamma \sin(2\omega_+ t)}{2\omega_+} \right), &\\
V_{22} \equiv& \langle p_1^2 \rangle = \frac{\cosh(2r)}{2} \left( \frac{J^2 - JK}{\omega_+^2} + \frac{4JK - 4K^2 - \Gamma^2}{4\omega_+^2} \cos(2\omega_+ t) + \frac{\Gamma \sin(2\omega_+ t)}{2\omega_+} \right), &\\
V_{33} \equiv& \langle q_2^2 \rangle = \frac{\cosh(2r)}{2} \left( \frac{J^2 + JK}{\omega_+^2} - \frac{4JK + 4K^2 + \Gamma^2}{4\omega_+^2} \cos(2\omega_+ t) - \frac{\Gamma \sin(2\omega_+ t)}{2\omega_+} \right), &\\
V_{44} \equiv& \langle p_2^2 \rangle = \frac{\cosh(2r)}{2} \left( \frac{J^2 - JK}{\omega_+^2} + \frac{4JK - 4K^2 - \Gamma^2}{4\omega_+^2} \cos(2\omega_+ t) - \frac{\Gamma \sin(2\omega_+ t)}{2\omega_+} \right), &\\
V_{12} =& V_{21} \equiv \left\langle q_1 p_1 + p_1 q_1 \right\rangle /2 = \frac{\sinh(2r)}{2} \left( -\frac{J \Gamma}{2\omega_+^2} + \frac{J \Gamma}{2\omega_+^2} \cos(2\omega_+ t) - \frac{J \sin(2\omega_+ t)}{\omega_+} \right), &\\
V_{13} =& V_{31} \equiv \left\langle q_1 q_2 + q_2 q_1 \right\rangle /2 = \frac{\sinh(2r)}{2} \left( \frac{4JK + 4K^2 + \Gamma^2}{4\omega_+^2} - \frac{J^2 + JK}{\omega_+^2} \cos(2\omega_+ t) \right), &\\
V_{14} =& V_{41} \equiv \left\langle q_1 p_2 + p_2 q_1 \right\rangle /2 = \frac{\cosh(2r)}{2} \left( \frac{J \Gamma}{2\omega_+^2} - \frac{J \Gamma}{2\omega_+^2} \cos(2\omega_+ t) - \frac{K \sin(2\omega_+ t)}{\omega_+} \right), &\\
V_{23} =& V_{32} \equiv \left\langle q_2 p_1 + p_1 q_2 \right\rangle /2 = \frac{\cosh(2r)}{2} \left( -\frac{J \Gamma}{2\omega_+^2} + \frac{J \Gamma}{2\omega_+^2} \cos(2\omega_+ t) - \frac{K \sin(2\omega_+ t)}{\omega_+} \right), &\\
V_{24} =& V_{42} \equiv \left\langle p_2 p_1 + p_1 p_2 \right\rangle /2 = \frac{\sinh(2r)}{2} \left( \frac{4JK - 4K^2 - \Gamma^2}{4\omega_+^2} + \frac{J^2 - JK}{\omega_+^2} \cos(2\omega_+ t) \right), &\\
V_{34} =& V_{43} \equiv \left\langle q_2 p_2 + p_2 q_2 \right\rangle /2 = \frac{\sinh(2r)}{2} \left( \frac{J \Gamma}{2\omega_+^2} - \frac{J \Gamma}{2\omega_+^2} \cos(2\omega_+ t) - \frac{J \sin(2\omega_+ t)}{\omega_+} \right). &
\end{align}
\end{subequations}

These have the same form as the variances in \cite{chakraborty2019delayed} if there is no two-mode squeezing interaction, i.e. $K=0$. Using the same notation as \cite{chakraborty2019delayed} we can write $V$ in the form
\begin{equation}
V = \begin{pmatrix}
    V_G & V_{GL}\\
     V^T_{GL} & V_L\\
\end{pmatrix},
\end{equation}
where $V_G$, $V_L$, and $V_{GL}$ are 2 × 2 sub-block matrices of $V$. The degree of quantum entanglement can be conveniently calculated by the logarithmic negativity $E_\text{N}$~\cite{plenio2005logarithmic,vidal2002computable}, which is given by
\begin{equation}
E_\text{N} = \max\left[0, -\ln 2\nu_- \right]
\end{equation}
with
\begin{equation}
\nu_-\equiv \frac{1}{\sqrt{2}}  \left[ \sum(V) - \sqrt{\sum(V)^2 - 4 \det(V)} \right]^{1/2},
\end{equation}
\begin{equation}
\sum(V) \equiv \det(V_G) + \det(V_L) - 2 \det(V_{GL}).
\end{equation}

\begin{figure}
    \centering
    \includegraphics[width=1\linewidth]{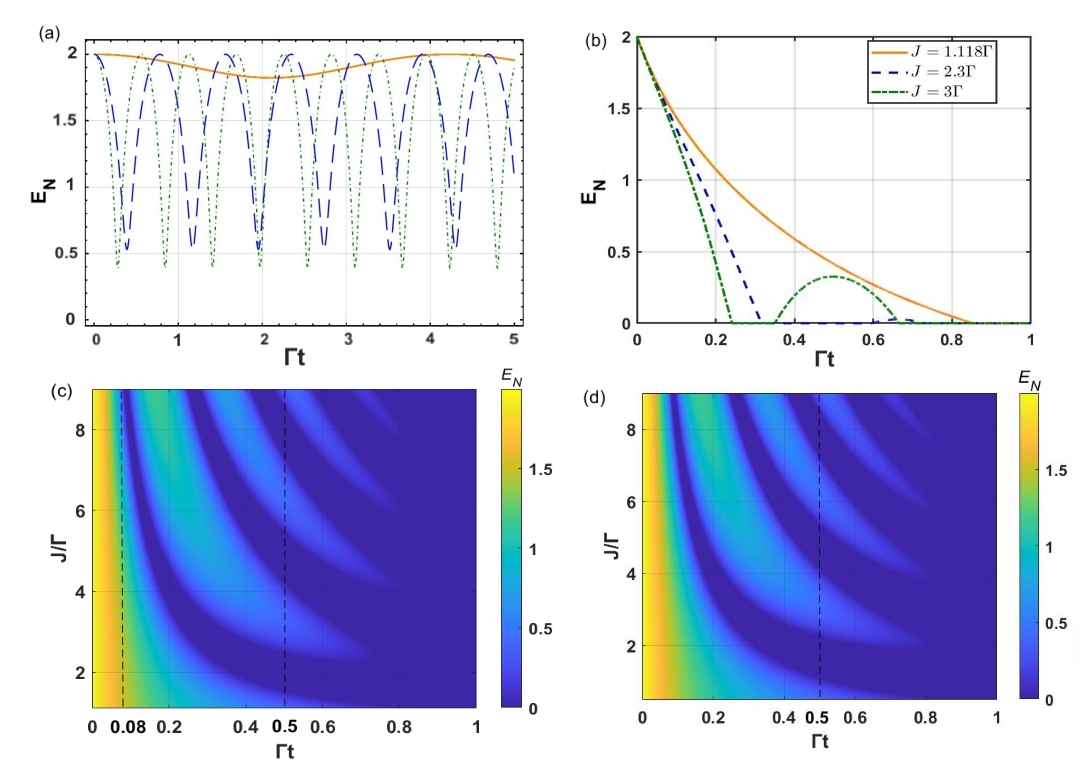}

    \caption{Dynamics of entanglement between the gain and loss resonators with parameters  \(r=1\) and \( \gamma=10^{-3} \). (a) Time evolution of entanglement in the absence of noise for \( K/\Gamma=1 \). The yellow (solid), blue (dashed), and green (dash-dotted) lines respectively correspond to \( J = 1.118\Gamma, 2.3\Gamma ,\) and  \( 3\Gamma \). (b) The same plot as (a) but with noise included for $n_{\text{th}}=10$. (c) Entanglement as a function of \( J/\Gamma \) and the time \(\Gamma t\) for \( K/\Gamma=1 \) and \( n_{\text{th}} = 10 \) . (d) Entanglement as a function of \( J/\Gamma \) and the time \(\Gamma t\) for  \( K/\Gamma=0 \)   and \( n_{\text{th}} = 10 \).}
    \label{fig:fig3}
\end{figure}
\begin{figure}
    \centering
     \includegraphics[width=1\linewidth]{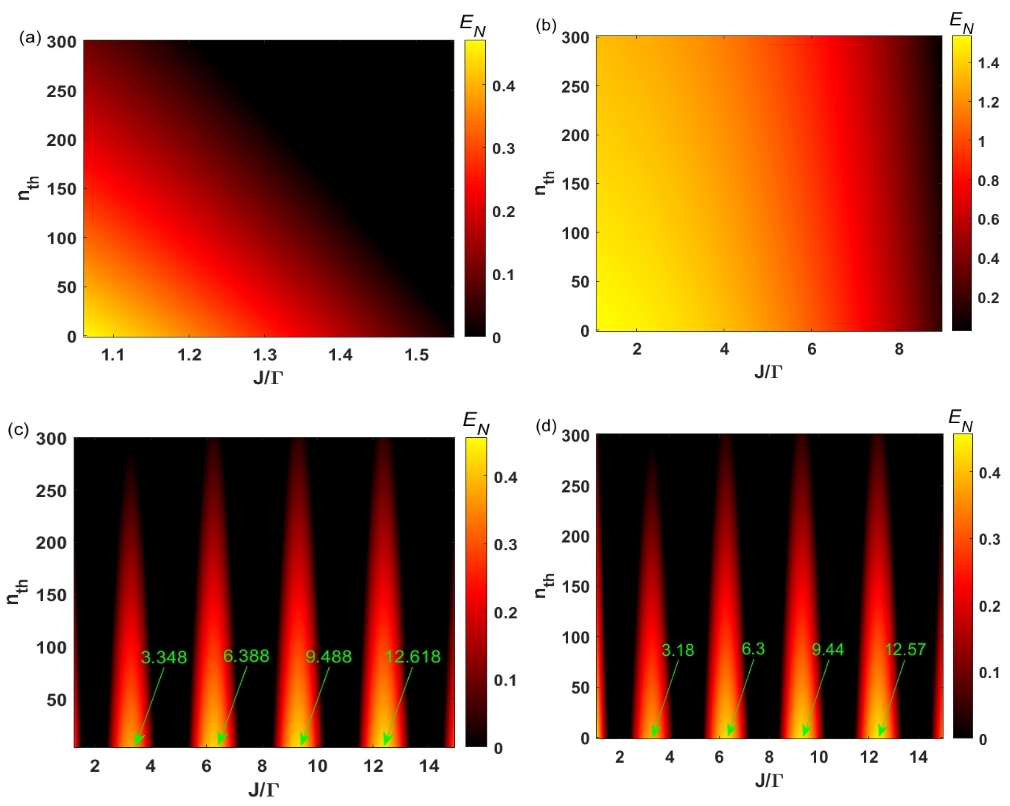}

        \caption{The robustness of entanglement in both pseudo-Hermitian ($K/\Gamma =1$) in (a)-(c) and $\mathcal{PT}$-symmetric ($K/\Gamma =0$) in (d) systems with parameters \(r=1\) and \( \gamma=10^{-3} \). (a) Entanglement at \( \Gamma t = 0.5 \) as a function of \( J/\Gamma \) and $n_{\text{th}}$ for \( K/\Gamma =1\) . (b) The same plot as (a) but with time $\Gamma t = 0.08$. (c) The same plot as (a) but with the extended parameter \( J/\Gamma \)  ranges from 1.118 to 15. (d) The same plot as (a) but with \( K/\Gamma =0\) and the extended parameter \( J/\Gamma \)  ranges from 0.5 to 15.}
    \label{fig:fig4}
\end{figure}
The dynamics of entanglement between the gain and the loss resonators is shown in Fig.~\ref{fig:fig3}. From Fig.~\ref{fig:fig3}(a) it can be seen that, in the absence of noise, the quantum entanglement exhibits periodic oscillations. However, the oscillations becomes less apparent (with lower amplitude and longer period) as the system approaches the EPs (i.e. \( J = 1.118 \Gamma \)). The results are similar to those in a $\mathcal{PT}$-symmetric system~\cite{chakraborty2019delayed}.
When taking the quantum noises into account, the situation is quite different. As evident in  Fig.~\ref{fig:fig3}(b), the entanglement exhibits both ESD and entanglement sudden birth (ESB) in the pseudo-Hermitian symmetric phase when \( J > 1.118 \Gamma \). Nevertheless, these phenomena decrease as $J/\Gamma$ approaches the EPs and when \( J = 1.118\Gamma \) the entanglement smoothly decreases with time, reaching zero at around \( \Gamma t \approx 0.842 \). Whereas, for \( J = 2.3\Gamma \) and \( J = 3\Gamma \), ESD occurs around \(\Gamma t \approx 0.316 \) and \(\Gamma t \approx 0.241 \) respectively. These cases also exhibit ESB at around \( \Gamma t \approx 0.597 \) and \( \Gamma t \approx 0.350 \) for \( J = 2.3\Gamma \) and \( J = 3\Gamma \) respectively. Despite ESB, it can be seen that entanglement is preserved for longer when our system is close to the EPs.
This shows that EPs in pseudo-Hermitian systems can have a similar effects to those in $\mathcal{PT}$-symmetric systems in terms of delaying ESD. However, we see that the introduction of two-mode squeezing interactions does not guarantee the preservation of the initial entanglement.  A reasonable explanation is that the two-mode squeezing interaction induces a squeezing operation in a specific direction within the phase space, thereby enhancing entanglement along that particular axis. However, the beam splitter interaction introduces a rotation in the phase space, effectively altering the orientation of the squeezed state. This rotation can misalign the squeezing direction, diminishing the entangling effect of the two-mode squeezing interaction. Consequently, the combined action of these two interactions may not lead to a net increase in entanglement, and in some cases, can even result in a reduction compared to the scenario where only the beam splitter interaction is applied. A fuller picture is shown in Fig.~\ref{fig:fig3}(c) where entanglement is plotted as a function of the coupling coefficient \( J/\Gamma \) and the time \(\Gamma t\), for \( n_{\text{th}} = 10 \). The plot shows that entanglement disappears more quickly and the value that reappears after ESB is more pronounced as the value of \( J/\Gamma \) increases. However, the time that entanglement completely disappears and never recovers is almost the same in each case, e.g., when \( J/\Gamma =8\), entanglement completely disappears at around \( \Gamma t \approx 0.83 \), which is slightly less than the time when we are at the EPs (\( \Gamma t \approx 0.842 \)). For comparison, in Fig.~\ref{fig:fig3}(d) we also plot entanglement as a function of \( J/\Gamma \) and the time \(\Gamma t\) for $\mathcal{PT}$-symmetric system by choosing \( K/\Gamma=0 \). The overall results are very similar for the pseudo-Hermitian and $\mathcal{PT}$-symmetric systems. The phenomenon of ESB in $\mathcal{PT}$-symmetric systems was not noted in~\cite{chakraborty2019delayed}, and we now turn our attention to exploring that in more detail.

The robustness of entanglement in both pseudo-Hermitian and $\mathcal{PT}$-symmetric systems is shown in Fig.~\ref{fig:fig4}. As indicated by the vertical dotted line in Fig.~\ref{fig:fig3}(c) and (d),  data are selected at two different times,  $\Gamma t = 0.5$ and  $\Gamma t = 0.08$, to illustrate our results.  As expected, there is a degradation of the entanglement with the increase of thermal phonons, however this is modest, suggesting that the entanglement is quite robust to thermal noise. Due to the effect of ESD, it is true that the closer we are to the EPs, the more robust the entanglement is to noise (see  Fig.~\ref{fig:fig4}(a) and also Fig.~8(a) of Ref.~\cite{chakraborty2019delayed}). However, when the parameter \( J/\Gamma \) is extended,  entanglement is seen to be very robust in the short time limit, as shown with $\Gamma t = 0.08$ in Fig.~\ref{fig:fig4}(b). At longer times, entanglement can be recovered for certain values of \( J/\Gamma \) as shown with  $\Gamma t = 0.5$ in Fig.~\ref{fig:fig4}(c). The maximum value of the recovered entanglement is almost the same as that near the EPs for the corresponding value of \( \Gamma t\).  This holds for both pseudo-Hermitian and $\mathcal{PT}$-symmetric systems (see Fig.~\ref{fig:fig4}(d)). From Eq.~(\ref{eigenfrequencies}), it can be calculated that the corresponding eigenfrequency difference for these certain values of \( J/\Gamma \)  is approximately $\pi$. In principle, for a given \(\Gamma t\), we can always find a suitable parameter far away from the EPs so that the entanglement is almost the same as the corresponding value at the EPs. This raises interesting questions about the cause of the beneficial effects seen in non-Hermitian systems and whether similar beneficial effects can be achieved at other parameters away from EPs. If so, this might be important in making these systems more experimentally accessible. Actually, a recent work has theoretically proposed and experimentally demonstrated universal non-Hermitian sensing in the absence of EPs\cite{xiao2024non}. This result, along with ours, opens a new perspective for the development of quantum devices of non-Hermitian systems in parameters even away from EPs.

\begin{figure}
    \centering
    \includegraphics[width=1\linewidth]{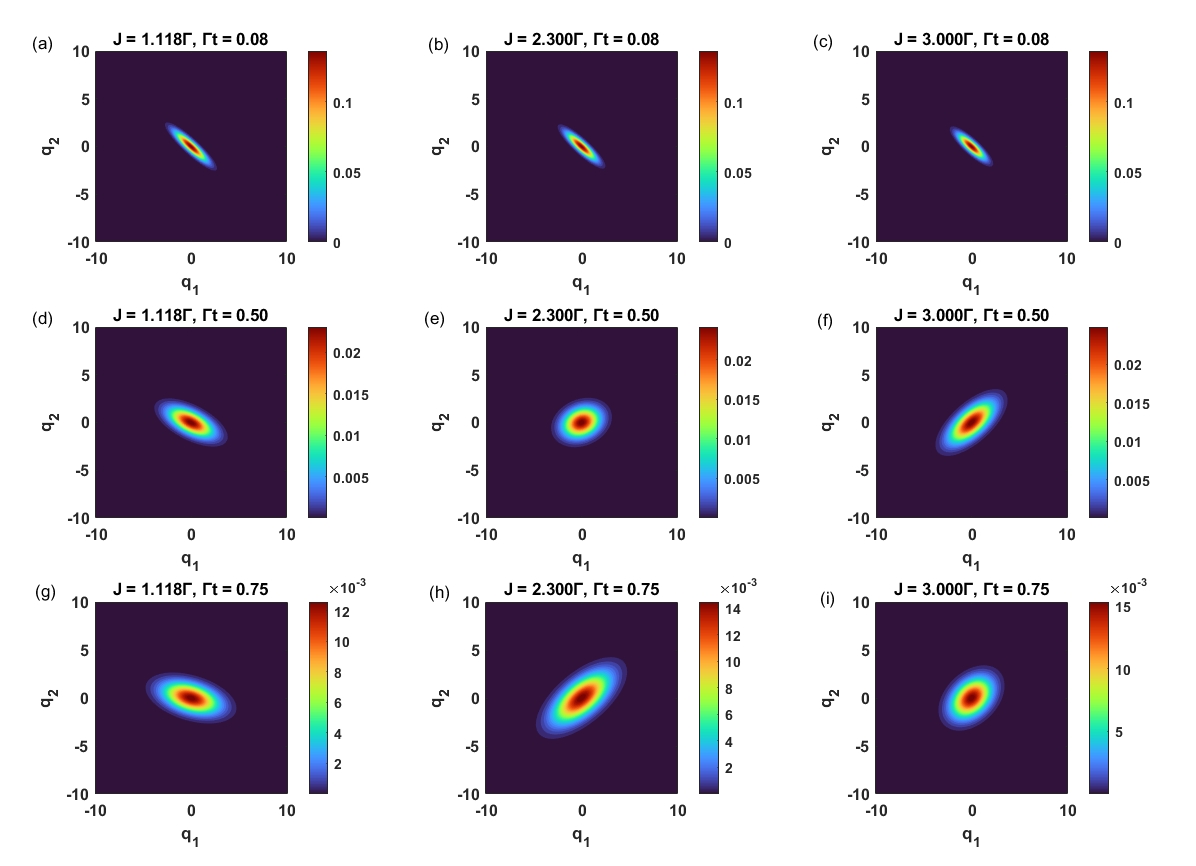}

        \caption{Time evolution of Wigner functions $W(q_1,q_2)_{p_2=0}^{p_1=0}$ for \( J = 1.118\Gamma,
  2.3\Gamma ,\) and  \( 3\Gamma \), respectively, at $\Gamma t=0.08$ (top three panels), $\Gamma t=0.5$ (middle three panels), and $\Gamma t=0.75$ (bottom three panels). Parameters common to each subplots are \(r=1\), \( \gamma=10^{-3}\), \(K/\Gamma=1\), and \( n_{\text{th}} = 10 \).}
    \label{fig:fig5}
\end{figure}
To further substantiate our claim, we analyzed the two-mode Wigner functions~\cite{adesso2007entanglement} at various times and coupling strengths, as shown in Fig.~\ref{fig:fig5}.

The Wigner function is defined by
\begin{align}
W(\mathbf{x}) = \frac{1}{\pi^2 \sqrt{\det V}} \exp\left(-\mathbf{x}^T V^{-1} \mathbf{x}\right),
\end{align}
where \( \mathbf{x} = ( q_1, p_1, q_2, p_2 )^T \).
As illustrated in Figs.~\ref{fig:fig5}(a)-(c),
at time \( t = 0.08 \), we observe squeezing behavior for coupling strengths \( J = 1.118\Gamma \), \( 2.3\Gamma \), and \( 3\Gamma \). This indicates the initial considerable amount of entanglement. At \( t = 0.5 \), the most pronounced squeezing is observed for \( J = 1.118\Gamma \), followed by \( J = 3\Gamma \), while \( J = 2.3\Gamma \) exhibits no significant squeezing, as illustrated in Fig.~\ref{fig:fig5}(d), ~\ref{fig:fig5}(f), and~\ref{fig:fig5}(e) respectively. At a later time \( t = 0.75 \), the strongest squeezing persists for \( J = 1.118\Gamma \). Notably,  \( J = 2.3\Gamma \) experiences a reemergence of squeezing, suggesting a revival of entanglement, whereas \( J = 3\Gamma \) display minimal squeezing, as depicted in Fig.~\ref{fig:fig5}(g), ~\ref{fig:fig5}(h), and~\ref{fig:fig5}(i). The above results are in good agreement with those in Figures 3 and 4, and support our conclusion.

To exhibit period behavior in an experiment, $J/K$ would need to be varied over quite a large range, which could be achieved as follows. The optomechanically-induced gain $\Gamma = 4G^2/\kappa=4g^2|\alpha|^2/\kappa$ can be adjusted by varying the field amplitudes $|\alpha|$ or the cavity decay rate $\kappa$. Consequently, the conditions $J/\Gamma \gg1$ and $K/\Gamma =1$ can be satisfied in an experiment. For example, in Ref.~\cite{ockeloen2018stabilized}, the parameters are $\gamma/2 \pi\sim 107 $Hz, $\kappa /2\pi\sim1335$kHz, and $G\sim283$kHz. The optomechanically-induced gain is calculated as $\Gamma = 4G^2/\kappa=4g^2|\alpha|^2/\kappa\sim240$kHz. The effective coupling between two mechanical oscillators ($J$ and $K$) is normally of the same order of magnitude as the effective optomechanical coupling $G\sim283$kHz, and thus comparable to the optomechanically-induced gain $\Gamma$. Nevertheless, by enhancing the cavity decay rate $\kappa$ within the bad cavity regime, we can effectively diminish $\Gamma$. This allows us to fulfill the conditions $J/\Gamma \gg1$ and $K/\Gamma =1$ by concurrently reducing the coupling $K$. Moreover, when we continue to increase $J/\Gamma$ up to 200, the entanglement still exhibits periodic changes. Theoretically, there is no upper limit to the ratio $J/\Gamma$. However, in practical applications, it is not necessary to make $J/\Gamma$ so large.

\section{Entanglement in ternary systems}\label{sec:Entanglement in ternary systems}

\begin{figure}
    \centering
    \includegraphics[width=0.75\linewidth]{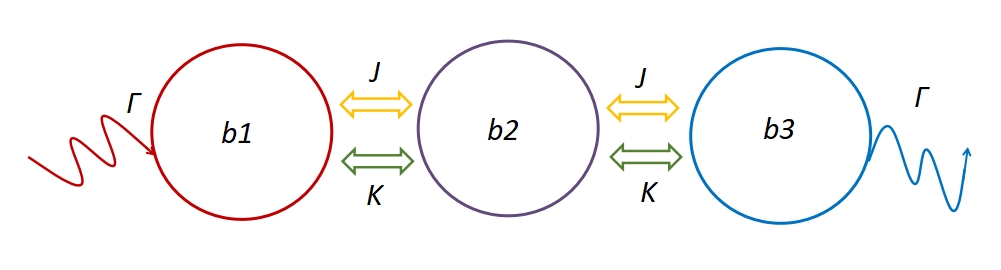}
    \caption{Schematic diagram of a ternary pseudo-Hermitian mechanical system consisting of a gain, neutral, and loss resonator respectively.}
    \label{fig:fig6}
\end{figure}

The above results can also be extended to a ternary system. According to Ref.~\cite{chakraborty2019delayed}, the equations governing a ternary mechanical system under $\mathcal{PT}$-symmetry are
\begin{subequations}
\begin{align}
\dot{b}_1 &= \left( \frac{\Gamma}{2} - \frac{\gamma}{2} \right) b_1 + iJ b_2 + i \sqrt{\Gamma} \, a_1^{\text{in}\dagger} + \sqrt{\gamma} \, b_1^{\text{in}}, \\
\dot{b}_2 &= -\frac{\gamma}{2} b_2 + iJ b_1 + iJ b_3 + \sqrt{\gamma} \, b_2^{\text{in}}, \\
\dot{b}_3 &= -\left( \frac{\Gamma}{2} + \frac{\gamma}{2} \right) b_3 + iJ b_2 + i \sqrt{\Gamma} \, a_2^{\text{in}} + \sqrt{\gamma} \, b_3^{\text{in}}.
\end{align}
\end{subequations}
Here, \( b_1 \), \( b_2 \), and \( b_3 \) (\( b_1^\dagger \), \( b_2^\dagger \), and \( b_3^\dagger \)) denote the annihilation (creation) operators of the gain, neutral, and loss resonators, respectively.
Following the approach presented in Section~\ref{sec:Entanglement in binary systems}, we construct a pseudo-Hermitian ternary mechanical system, in which the gain and loss oscillators are separated by a neutral resonator, as shown in Fig.~\ref{fig:fig6}. Therefore, each mechanical resonator satisfies the following equations of motion
\begin{subequations}
\begin{align}
\dot{b}_1 &= \left( \frac{\Gamma}{2} - \frac{\gamma}{2} \right) b_1 + iJ b_2 - iK b_2^{\dagger}+ i \sqrt{\Gamma} \, a_1^{\text{in}\dagger} + \sqrt{\gamma} \, b_1^{\text{in}}, \\
\dot{b}_2 &= -\frac{\gamma}{2} b_2 + iJ b_1- iK b_1^{\dagger}+ iJ b_3- iK b_3^{\dagger} +  \sqrt{\gamma} \, b_2^{\text{in}}, \\
\dot{b}_3 &= -\left( \frac{\Gamma}{2} + \frac{\gamma}{2} \right) b_3 + iJ b_2 - iK b_2^{\dagger}+ i \sqrt{\Gamma} \, a_2^{\text{in}} + \sqrt{\gamma} \, b_3^{\text{in}}.
\end{align}
\end{subequations}

\begin{figure}
    \centering
    \includegraphics[width=1\linewidth]{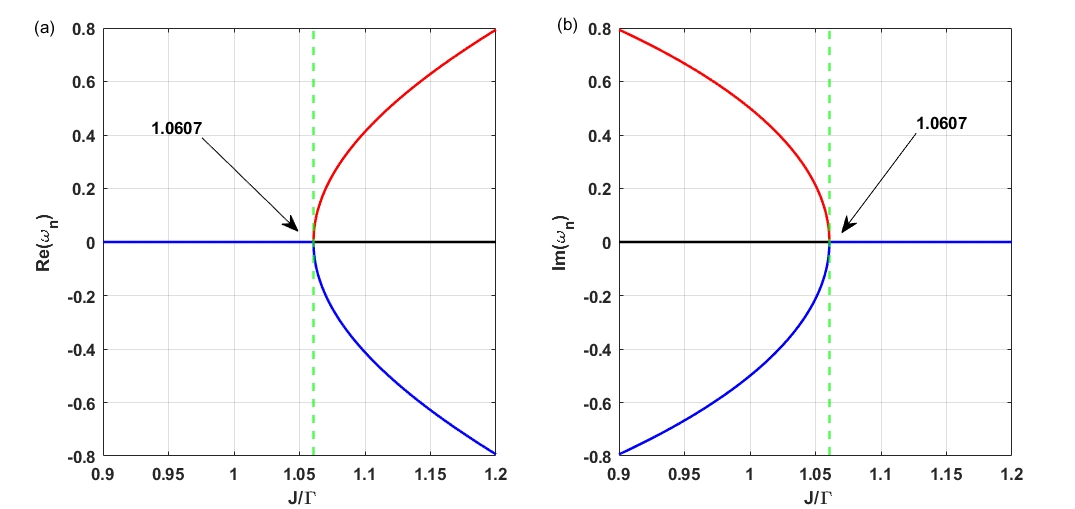}

    \caption{(a) The real and (b) imaginary parts of the eigenfrequencies \( \omega_\pm \) as a function of the ratio of coupling coefficients \( J/\Gamma \) when \( K/\Gamma  = 1 \). The exceptional point appears at \( J/\Gamma = 1.0607 \).}
    \label{fig:fig7}
\end{figure}

Similar to the binary system, we obtain a cubic algebraic equation in the absence of any noise, and each eigenfrequency \( \omega_n \) (\( n = -1, 0, 1 \)) satisfies
\begin{equation}\label{algebraic}
\omega_n \left( \omega_n^2 + \frac{\Gamma^2}{4} - 2J^2+ 2K^2\right) = 0.
\end{equation}
According to Eq.~(\ref{algebraic}), the critical coupling strength at which the three eigenfrequencies coalesce is \( J/\Gamma = 1.0607 \). This feature is illustrated in Fig.~\ref{fig:fig7}.
\begin{figure}
    \centering
    \includegraphics[width=1\linewidth]{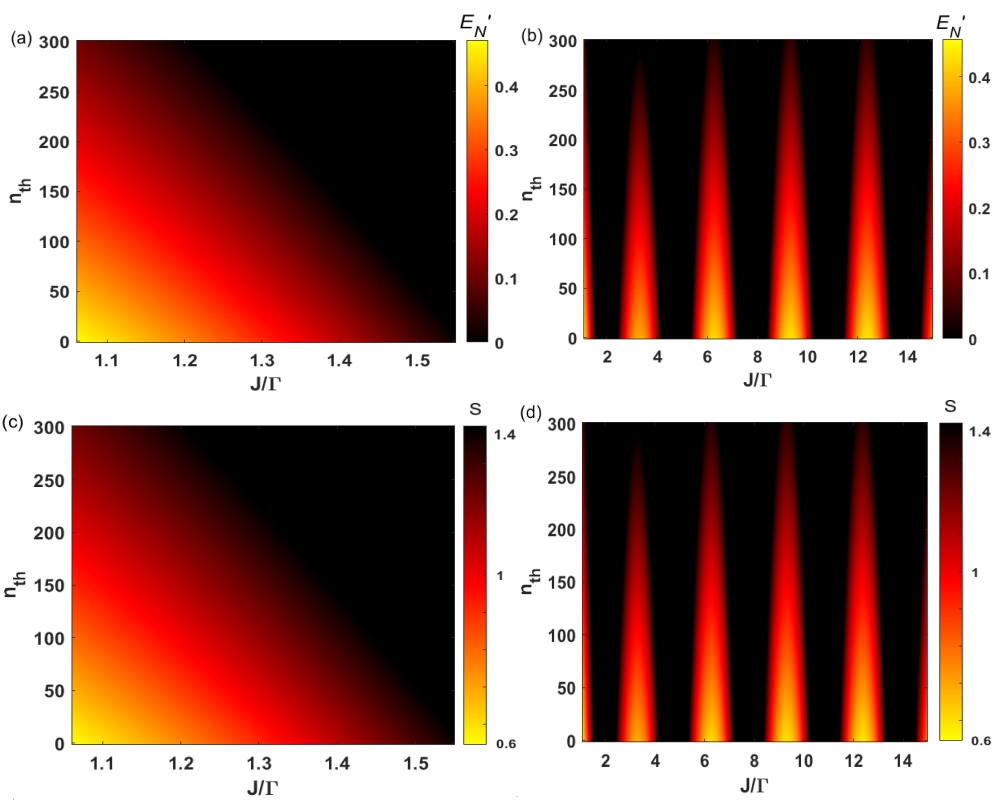}

    \caption{The robustness of entanglement and nonseparability criterion $S$ in our pseudo-Hermitian ($K/\Gamma = 1$) system with parameters \( r = 1 \) and \( \gamma = 10^{-3} \). (a) Entanglement at \( \Gamma t = 0.5 \) as a function of \( J/\Gamma \) and \( n_{\text{th}} \) . (b) The same plot as (a) but with the extended parameter \( J/\Gamma \) ranging from 1.0607 to 15. (c) Nonseparability criterion $S$ at \( \Gamma t = 0.5 \) as a function of \( J/\Gamma \) and \( n_{\text{th}} \) . (d) The same plot as (c) but with the extended parameter \( J/\Gamma \) ranging from 1.0607 to 15.}
    \label{fig:fig8}
\end{figure}

To quantify entanglement in the ternary system, we first adopt logarithmic negativity as the entanglement measure, calculating the entanglement between any single mode and the remaining two modes $E_N'$~\cite{adesso2006multipartite}. To do this, a partial transposition operation is performed on the selected mode, and the minimum symplectic eigenvalue $\tilde{\nu}_{\min}$ is obtained by constructing the standard symplectic matrix
\begin{equation}
\Omega = \bigoplus_{j=1}^{n}
\begin{bmatrix}
0 & 1 \\
-1 & 0
\end{bmatrix}.
\end{equation}
The logarithmic negativity between one mechanical oscillator and the remaining two is then defined as
\begin{equation}
E_N' = \max\left[ 0, -\ln(2\tilde{\nu}_{\min}) \right],
\end{equation}
where $\tilde{\nu}_{\min}$ denotes the smallest symplectic eigenvalue of the partially transposed covariance matrix.
We also evaluate the inseparability criterion~\cite{van2003detecting,teh2014criteria,gonzalez2018continuous} by
defining the following linear combinations of orthogonal quadratures:
\begin{subequations}
\begin{align}
x &\equiv h_1 q_1 + h_2 q_2 + h_3 q_3, \\
y &\equiv g_1 p_1 + g_2 p_2 + g_3 p_3,
\end{align}
\end{subequations}
where \( q_1\), \(q_2\), and \(q_3 \) (\( p_1\), \(p_2\), and \(p_3 \)) are the position (momentum) operators of the gain, neutral, and loss resonators, respectively, while \( h_k \) and \( g_k \) are arbitrary real parameters. Then, according to~\cite{van2003detecting,teh2014criteria,gonzalez2018continuous}, the inseparability criterion is given by
\begin{align}
S = \expval{(\Delta x)^2} + \expval{(\Delta y)^2}.
\end{align}

A tripartite quantum state is considered genuinely tripartite entangled if and only if it violates the following single inequality:
\begin{equation}
S \geq \min
\big\{
|h_{3}g_{3}| + |h_{1}g_{1} + h_{2}g_{2}|,
|h_{2}g_{2}| + |h_{1}g_{1} + h_{3}g_{3}|,
|h_{1}g_{1}| + |h_{2}g_{2} + h_{3}g_{3}|
 \big\},
\end{equation}
whereas only the violation of any of these inequalities
\begin{equation}
S \geqslant \big( |h_k g_k| + |h_l g_l + h_m g_m| \big),
\end{equation}
for a given permutation of \( k, l, m \) of \( 1, 2, 3 \), ensures a full tripartite inseparability.
Similar to~\cite{chakraborty2019delayed}, we numerically calculate the lower bound of \( S \)
 and identify the conditions under which genuine tripartite entanglement can occur by constructing the following set of parameters
\begin{align}
&h_1 = g_1 = 1, &g_2 = g_3 = -h_2 = -h_3 = \tfrac{1}{\sqrt{2}}.
\end{align}

Then, to ensure genuine tripartite nonseparability, the condition \( S < 1 \) must be satisfied, while \( S < 2 \) is sufficient to confirm at least complete tripartite nonseparability.

From Fig.~\ref{fig:fig8}(a), it is evident that the entanglement between one mechanical oscillator and the remaining two reaches its maximum at $J = 1.0607$. Subsequently, this entanglement gradually decays as the thermal phonon number $n_{\mathrm{th}}$ and the coupling coefficient \( J/\Gamma \) increase, exhibiting behavior analogous to that observed in binary systems. When the coupling coefficient \( J/\Gamma \) is extended, the entanglement displays periodic oscillations, as illustrated in Fig.~\ref{fig:fig8}(b).
As depicted in Fig.~\ref{fig:fig8}(c), the tripartite entanglement demonstrates considerable robustness near the EPs. However, as the system deviates from this point, the entanglement gradually decreases and, similar to the behavior in a binary system, exhibits periodic oscillations over \( J/\Gamma \), as shown in Fig.~\ref{fig:fig8}(d). This further confirms the robustness of entanglement even away from EPs.

\section{Conclusions}\label{sec:Conclusions}
In conclusion, we have extended previous studies of $\mathcal{PT}$-symmetric cavity-optomechanical systems in two key ways. We have investigated their dynamics far from EPs and also extended them into the pseudo-Hermitian regime by introducing a two-mode squeezing interaction into a system that was already characterised by $\mathcal{PT}$-symmetry. In both cases we found similar behaviors including entanglement sudden death and entanglement sudden birth. These phenomena originate from the systems' non-Hermitian character and the specific parameter selections, where EPs and nonlinear interactions play an important role. We also observed that the entanglement is remarkably robust to thermal noise even when we are far from an EPs, which may prove useful for practical applications. Our research advances our understanding of entanglement phenomena in non-Hermitian systems and establishes theoretical foundations for the design of quantum devices with specialized capabilities both near and far from EPs.

\section*{Acknowledgments}
 This work is supported by the National Natural Science Foundation of China (Grants No.~12004336, No.~12075205, and No.~62071430), the International Exchanges 2022 Cost Share (NSFC) with ref: IEC\verb|\|NSFC\verb|\|223131, the Fundamental Research Funds for the Provincial University of Zhejiang (Grant No.~XRK23006), the Natural Science Foundation of Fujian Province (Grant No.~2020J05199), the Foundation of the Zhejiang Provincial Department of Education (Grant No.~Y202455809), the Dr. Start Funding from Longyan University (LB2020005), and Funds from the China Scholarship Council.

\bibliography{references}
\bibliographystyle{apsrev4-1}

\end{document}